\documentclass{article}

\usepackage[english]{babel}

\usepackage[letterpaper,top=2cm,bottom=2cm,left=3cm,right=3cm,marginparwidth=1.75cm]{geometry}

\usepackage{amsmath}

\usepackage{graphicx}
\usepackage[colorlinks=true, allcolors=blue]{hyperref}

\title{Many-Eyes and Sentinels in Selfish and Cooperative Groups}

\author{Charlie Pilgrim$^{1,*}$ \and 
    Andrew M Bate$^{1}$ \and 
    Anna Sigalou$^{2}$ \and
    Mélisande Aellen$^{3}$ \and
    Joe Morford$^{3}$ \and
    Elizabeth Warren$^{4}$ \and
    Christopher Krupenye$^{4}$ \and
    Dora Biro$^{3}$ \and
    Richard P Mann$^{1}$}

\date{}

\begin{document}
\maketitle

\noindent
$^{1}$School of Mathematics, University of Leeds, Leeds, UK\\
$^{2}$Consejo Superior de Investigaciones Científicas, Spain \\
$^{3}$Department of Brain and Cognitive Sciences, University of Rochester, Rochester, USA\\
$^{4}$Department of Psychological \& Brain Sciences, Johns Hopkins University, Baltimore, USA\\
\\
$^{*}$Corresponding author: c.p.pilgrim@leeds.ac.uk

\begin{abstract}

Collective vigilance describes how animals in groups benefit from the predator detection efforts of others. Empirical observations typically find either a many-eyes strategy with all (or many) group members maintaining a low level of individual vigilance, or a sentinel strategy with one (or a few) individuals maintaining a high level of individual vigilance while others do not. With a general analytical treatment that makes minimal assumptions, we show that these two strategies are alternate solutions to the same adaptive problem of balancing the costs of predation and vigilance. Which strategy is preferred depends on how costs scale with the level of individual vigilance: many-eyes strategies are preferred where costs of vigilance rise gently at low levels but become steeper at higher levels (convex; e.g. an open field); sentinel strategies are preferred where costs of vigilance rise steeply at low levels and then flatten out (concave; e.g. environments with vantage points). This same dichotomy emerges whether individuals act selfishly to optimise their own fitness or cooperatively to optimise group fitness. The model is extended to explain discrete behavioural switching between strategies and differential levels of vigilance such as edge effects.

\end{abstract}

\subsubsection{Keywords}

collective vigilance | many-eyes | sentinels | collective intelligence | collective sensing | cooperation | game theory

\section{Introduction}



Animals living in groups benefit from the vigilance efforts of others, whereby if one individual spots a predator approach then others in the group are alerted \cite{Pulliam1973Feb, beauchamp2015animal}. However, not all groups use the same strategy to achieve these benefits. In some species we see ``many-eyes" behaviour, where numerous individuals keep a low level lookout to collectively generate a high chance of detecting predators \cite{powell1974experimental, fernandez2007group, elgar1989predator, kenward1978hawks, siegfried1975flocking, hoogland1979effect, monaghan1985group, li2009vigilance, marino2008vigilance, magurran1985vigilant}. In other cases we see sentinel behaviour, where one or a few individuals maintain high vigilance while others forage  \cite{wickler1985coordination, gaston1977social, beauchamp2022visual, clutton1999selfish, huels2022sentinel, manser1999response, rasa1986coordinated, Horrocks_1986, fox2014rabbitfish}. Despite decades of theoretical and empirical research, we lack a unified framework explaining the ecological conditions that favour each strategy \cite{bednekoff2015sentinel}. 

Many-eyes behaviour is widespread and has been observed in varied taxa spanning birds \cite{powell1974experimental, fernandez2007group, elgar1989predator, kenward1978hawks, siegfried1975flocking}, mammals \cite{hoogland1979effect, monaghan1985group, li2009vigilance, elgar1989predator, marino2008vigilance}, fish \cite{magurran1985vigilant}, and insects \cite{treherne1981group}. Enhanced predator detection is considered a key benefit of living in groups, with the``many-eyes hypothesis" \cite{Pulliam1973Feb, treisman1975predation, dimond1974problem} predicting empirical patterns that larger groups enjoy a greater collective chance of detecting predators \cite{powell1974experimental, kenward1978hawks, siegfried1975flocking, marino2008vigilance, magurran1985vigilant} while incurring lower individual costs of vigilance \cite{hoogland1979effect, monaghan1985group, li2009vigilance, powell1974experimental, fernandez2007group, elgar1989predator}.  A canonical example is the yellow-eyed junco (\emph{Junco phaeonotus}): these birds forage in groups on the ground for seeds while occasionally scanning for predators such as hawks, raising alarm calls upon detection \cite{pulliam1982scanning, caraco1979time}.  

Sentinel behaviour, while not as prevalent as many-eyes behaviour, is taxonomically widespread and observed in birds \cite{wickler1985coordination, gaston1977social, beauchamp2022visual}, mammals \cite{clutton1999selfish, huels2022sentinel, manser1999response, rasa1986coordinated, Horrocks_1986}, fish \cite{fox2014rabbitfish}, and insects \cite{shackleton2018organization}. Sentinels typically take up an elevated position \cite{wickler1985coordination, clutton1999selfish, gaston1977social, rasa1986coordinated, Horrocks_1986, bednekoff2001coordination, olson2015exploring, bednekoff2015sentinel} and produce alarm calls upon detecting threats \cite{clutton1999predation, gaston1977social, rasa1986coordinated, bednekoff2001coordination}. A canonical example is meerkats (\emph{Suricata suricatta}), where sentinels stand upright on a mound or tree while the rest of the group forage for food, with roles switching throughout the day \cite{clutton1999selfish, manser1999response, santema2013meerkat}.  Conditions that are thought to support sentinel behaviour include habitats with good vantage points \cite{rasa1986coordinated, bednekoff2001coordination, bednekoff2015sentinel}, the ability to coordinate \cite{wickler1985coordination, gaston1977social, rasa1986coordinated, ward1985birds, bednekoff2015sentinel}, safe sentinel positions \cite{bednekoff1997mutualism, bednekoff2001coordination, pulliam1982scanning}, mechanisms for cooperation \cite{pulliam1982scanning, wickler1985coordination, Horrocks_1986, ridley2013sentinel}, high predation risk \cite{rasa1986coordinated, ward1985birds, bednekoff2001coordination, clutton1999selfish}, and resource abundance \cite{gaston1977social, clutton1999selfish, bednekoff2015sentinel}.

Collective vigilance creates a social dilemma: the benefits of predator detection are shared throughout the group while costs are paid individually, generating incentives to free-ride by reducing one's own individual vigilance while benefiting from the vigilance efforts of others \cite{pulliam1982scanning}. Game-theoretic models demonstrate that many-eyes behaviour can arise in selfish groups, yet empirical observations have been interpreted as either selfish or cooperative depending on the underlying assumptions \cite{pulliam1982scanning, mcnamara1992evolutionarily, packer1990should}. Early work on sentinel behaviour interpreted it as cooperative within a kin-selection framework \cite{rasa1987vigilance, mcgowan1989sentinel}, but later models showed that sentinels can be favoured by pure self-interest if sentinel positions are safer than foraging \cite{bednekoff2001coordination, bednekoff1997mutualism}. Empirical evidence is mixed here too, with some support for ``safe selfish sentinels" \cite{clutton1999selfish}, while other work finds evidence for cooperative dynamics \cite{Horrocks_1986, ridley2013sentinel, rasa1986coordinated, rasa1987vigilance, bednekoff2015sentinel, santema2013meerkat}.

Existing theoretical work has typically treated many-eyes and sentinel behaviours as distinct phenomena. Influential many-eyes models focus on how vigilance changes with group size and cooperation \cite{Pulliam1973Feb, pulliam1982scanning, mcnamara1992evolutionarily, packer1990should}, while influential sentinel models focus on coordination, cooperation, and the safety of sentinel positions \cite{bednekoff1997mutualism, bednekoff2001coordination, wickler1985coordination}. Each framework captures only a subset of ecological conditions and behaviour with fixed behavioural rules or specific cost-benefit functions. Consequently, the theoretical literature is fragmented and resists synthesis, with a recent review highlighting ``large gaps in knowledge" \cite{bednekoff2015sentinel} in the social and environmental conditions favouring sentinel over many-eyes strategies. 

Here we present a general model that reveals that many-eyes and sentinel strategies are alternate solutions to the same underlying fitness optimisation problem. Which is preferred depends on how the individual costs of vigilance scale. In uniform habitats (e.g. open fields), vigilance costs are \emph{convex}: low levels of vigilance are cheap but costs are steepening as individuals sacrifice more foraging time at higher levels of vigilance, favouring a strategy of distributing vigilance throughout the many-eyes of the group. In structured habitats with vantage points (e.g. trees or mounds) vigilance costs are \emph{concave}: it costs a lot to reach a vantage point but increasing vigilance once there is cheap, favouring the concentration of vigilance in sentinels. This same dichotomy emerges in both selfish and cooperative groups. We demonstrate this with general analytical results, supported by simulations. We then extend the model to explain behavioural switching between strategies \cite{clutton1999selfish} and differential vigilance such as edge effects \cite{elgar1989predator, inglis1981vigilance, jennings1980influence}, before discussing implications for the convergent evolution of collective vigilance. 

\section{A General Model of Collective Vigilance}

The adaptive challenge of collective vigilance is balancing the threat of predation with the costs of vigilance \cite{caraco1979time, pulliam1982scanning, mcnamara1992evolutionarily, bednekoff2001coordination, bell2009value, van2022modeling, lima2021influence, olson2015exploring, lima1990behavioral}. Considering this challenge, we write down a general model of collective vigilance. The net fitness of individual $i$ is 

\begin{equation}
    f_i = b(S) - c(v_i) \,, \label{eqn: fitness}
\end{equation}

where $S$ is the collective vigilance, $b(S)$ is the benefit of predator detection, $v_i$ is the individual vigilance, and $c(v_i)$ is the individual cost of vigilance. Below, we define each of these terms clearly and show how this general model encompasses many existing models in the literature, including scanning models \cite{Pulliam1973Feb, pulliam1982scanning, delm1990vigilance}, probability models \cite{rubenstein1978predation, packer1990should, monaghan1985group, ward2011fast}, and hazard survival models \cite{mcnamara1992evolutionarily, gilliam1987habitat, houston1993general}. 

Throughout our analysis, we primarily work with this general model without relying on specific forms for the benefits and costs of vigilance. Analysing vigilance in this way means that our results are truly general, although some of the analytical treatment is necessarily abstract.  
 
\subsection{Defining Vigilance}

We consider individual vigilance, $v_i$, as a general abstract measure of vigilance that could represent a range of physical realities e.g. rate of scans for predators \cite{pulliam1982scanning, Pulliam1973Feb}, time spent scanning \cite{monaghan1985group, mcnamara1992evolutionarily, ward1985birds}, the probability of detecting a predator \cite{packer1990should, ward2011fast}. The definition of individual vigilance is also constrained by Equation \ref{eqn: fitness} such that the benefit, $b(S)$, is a function of the collective vigilance, which we define as an aggregate of the individual vigilances 

\begin{equation}
    S = \sum_i v_i \,. \label{eqn:collective_vigilance}
\end{equation}

This definition straightforwardly aligns with existing literature that defines individual vigilance, $v_i$, as the rate of scanning for predators \cite{Pulliam1973Feb, pulliam1982scanning, delm1990vigilance}. In these models, successful predator approaches require a time window, $\tau$, in which no individual scans (assuming that if any individual spots the predator approach then they will warn the rest of the group). For each individual, the probability of detecting an approaching predator is $p = 1 - e^{-v_i \tau}$. For a collective, the probability of any (one or more) individuals detecting an approaching predator is $p_C = 1 - e^{- S \tau }$, where the collective vigilance, $S$, is the sum of individual vigilances. If individual scans are independent Poisson processes with rate $v_i$, then the collective rate of scanning is described by a combined Poisson process with rate $S = \sum_i v_i$. 

In other literature, vigilance is defined more directly in terms of the probability of an individual detecting an approaching predator, $p$, so that a predator approach is successful if all of the individuals fail to detect the approach in time, given by $p_C = 1 - (1 - p)^n$ \cite{rubenstein1978predation, packer1990should, monaghan1985group, ward2011fast}. At first glance this seems to be incompatible with our definition in Equation \ref{eqn:collective_vigilance}. However, if we define the vigilance as $v_i = -ln(1-p)$, then the probability of the group detecting an approaching predator is a function of the collective vigilance, $p_C = 1 - e^{-S}$ (and the probability for a single individual is $p = 1 - e^{-v_i}$), similar to the scanning model. Note that defining vigilance in this way changes nothing about the actual system we are studying, but it does unite models that describe vigilance in terms of rates of scans for predators with those which more directly work with probabilities. 

Some models in the literature are multiplicative. For example, hazard models assume that fitness benefits accrue until an individual is predated \cite{gilliam1987habitat, houston1993general, mcnamara1992evolutionarily}. In a simple form, expected fitness can be written as $f_i = T \rho$, where $T$ is the expected survival time and $\rho$ is the rate of fitness gain. To accommodate such multiplicative models in our framework, one could log-transform these models to separate them into the general additive form in Equation \ref{eqn: fitness}, i.e. $\log(f_i) = \log(T) + \log(\rho)$, so that the benefit is $b(S) = \log(T)$ and the cost is $c(v_i) = - \log(\rho)$. The rest of our analysis then follows, since maximising the log-transformed fitness is equivalent to maximising the non-transformed function. 

The general model that we present is homogeneous in that every individual is the same in every way, although they can have different individual vigilances, $v_i$. The model also involves a mean-field approximation in that individuals experience collective vigilance through an aggregate of their individual-level vigilances, $S$. In reality, there are many sources of individual heterogeneity in groups \cite{hoogland1979effect, lima1990behavioral, clutton1999selfish, wright2001safe, olson2015exploring, van2022modeling, barnard1982time}. In addition, the benefit of vigilance is not purely shared and it can be more beneficial to personally spot a predator first \cite{bednekoff1997mutualism, bednekoff2001coordination, mcnamara1992evolutionarily, sirot2009coordination, bednekoff1998re, lima1990behavioral}. We discuss these realistic considerations in the Extensions section. 

\subsection{The Benefit of Collective Vigilance, $b(S)$}

Without defining a specific form for the benefit function, $b(S)$, we can nevertheless say something about its general shape. 

The benefit of collective vigilance is in reducing the threat of predation. We define the predation threat, $r$, as an extrinsic threat level (equivalent to the expected individual fitness cost of predation if the group was not vigilant at all). Fundamentally, the most one can do is to completely remove this threat, i.e. reduce the probability of predation to zero (or a minimum), in which case the benefit of collective vigilance would be $b(S) = r$. Formally, we can say that $0 \leq b(S) \leq r$. The benefit increases with collective vigilance, $\frac{d b}{d S} > 0$. However, because the maximum benefit is capped, benefits cannot increase indefinitely and they must eventually saturate. We capture this by assuming that marginal benefits are decreasing, $\frac{d ^2 b}{d S^2} < 0$, so that the benefit function is increasing and concave. 

The predation threat appears in existing literature \cite{ bednekoff2001coordination, bell2009value, lima1990behavioral}. In our definition, the single variable $r$ captures a range of considerations including the rate of predator approaches \cite{mcnamara1992evolutionarily, lima1990behavioral}, the chance of individual predation given a successful approach \cite{mcnamara1992evolutionarily, lima1990behavioral}, as well as the relative fitness cost of being predated (e.g. the predation threat can be expressed as a multiplication of these considerations). The level of predation threat is also influenced by other group anti-predator benefits such as dilution \cite{delm1990vigilance, powell1974experimental, treisman1975predation} and confusion \cite{powell1974experimental, kenward1978hawks, milinski1984predator} effects, which we assume are fixed for a given group size. 

\subsection{The Costs of Vigilance and Environmental Structure}

As with benefits, we can say something about the general shape of the cost of vigilance function, $c(v_i)$, without defining a specific form. 

Individuals pay a fitness cost for maintaining vigilance. In the literature this cost is often connected (explicitly or implicitly) to the opportunity costs of foraging \cite{caraco1979time, mcnamara1992evolutionarily, bednekoff2001coordination, bell2009value, van2022modeling, lima2021influence, olson2015exploring}. Time spent looking out for predators is time not spent foraging for food (or engaging in other fitness enhancing activities).  Note that in our model the costs of vigilance do not include predation costs, which are fully captured by the benefit function as described above. 

We assume that each individual can have a range of levels of vigilance, $v_i \in [0, v_{max}]$. Individual vigilance costs scale with the level of vigilance as described by a function $c(v_i)$. We assume that the cost function is increasing, $\frac{d c}{d v_i} > 0$, i.e. higher levels of vigilance incur higher costs. As fitness is relative, we arbitrarily set $c(0)=0$. 

Some habitats are approximately uniform in the sense that predator approaches can be observed equally well from any position \cite{bednekoff2015sentinel} (such as an open field). In this case, we can reasonably approximate the costs of vigilance as the opportunity cost of the time not spent foraging for food \cite{rasa1986coordinated}. Following the literature, we consider that fitness costs for a low level of vigilance are relatively small \cite{mcnamara1992evolutionarily, beauchamp2015animal, lima1999back}, as the time spent being vigilant has a small effect on foraging efforts. How do these costs scale when an individual spends more time being vigilant? At one extreme, if an individual spent all their time being vigilant and no time foraging then they would not survive, in which case their fitness cost would be very high. Following existing work \cite{mcnamara1992evolutionarily}, we assume that the costs of vigilance in uniform habitats start off shallow at low vigilance before becoming steeper at high vigilance, $\frac{d^2c}{d v_i^2} > 0$, i.e. costs of vigilance are steepening (convex). 

Other habitats have areas that have low visibility for detecting predators and other areas that provide better vantage points (such as mounds or trees) \cite{bednekoff2015sentinel, wickler1985coordination, clutton1999selfish, gaston1977social, rasa1986coordinated, Horrocks_1986, bednekoff2001coordination, olson2015exploring}. The cost of vigilance in this case starts off steep, as in order to increase vigilance the individual must remove themselves from foraging and move to the vantage point. However, once at the vantage point the cost of increasing vigilance is relatively cheap: once you are up the tree you may as well keep your eyes peeled. These environments are characterised by flattening (concave) costs of vigilance $\frac{d^2c}{d v_i^2} < 0$. 

Of course, vigilance costs can be more complicated and can have regions of convexity and concavity. We discuss these considerations in the Extensions section. 

\section{Results}

\subsection{Selfish Groups}

Selfish groups are made up of $N$ individuals who are incentivised to maximise their own individual fitness \cite{mcnamara1992evolutionarily}, where fitness is relative to a reference population larger than the group (Equation \ref{eqn: fitness}). Because the benefit of vigilance depends only on the level of collective vigilance, $S = \sum v_i$, all group members share the same benefit, $b(S)$, as well as the same marginal benefit $\frac{d b}{d S}$. Following existing work \cite{bednekoff1997mutualism} we model behaviour as a dynamic game whereby individuals adapt their behaviour to the ecological context and the vigilance levels of the rest of the group. 

Consider first an environment without vantage points. Individuals increase their vigilance at the opportunity cost of foraging, with steepening (convex) vigilance costs such that marginal costs increase with the individual level of vigilance. If Alice has higher vigilance than Bob, then she is paying a higher marginal cost but receiving the same benefit, so Bob is more incentivised than Alice to increase his vigilance. If the collective vigilance is high enough then Alice is incentivised to reduce her vigilance while Bob increases his. Their vigilances converge. Extending this argument to the entire group, the only stable outcome is for everyone to have the same vigilance, i.e. a many-eyes strategy. 

Now consider an environment with a vantage point, such as a tree or mound. Reaching the vantage point is costly, but once there maintaining and increasing vigilance are relatively cheap, so that costs are flattening (concave). If the predation threat is high enough then Alice finds it beneficial to pay the cost of climbing the vantage point to keep watch. Once there, Alice has low marginal costs while Bob (still on the ground) has high marginal costs. As they share the same marginal benefit, Alice is incentivised to increase her vigilance while Bob reduces his to zero. This divergence generates a sentinel strategy with one or a few individuals maintaining high vigilance while the rest of the group have zero vigilance. 

More formally, each individual's incentive to change vigilance is determined by the slope of their fitness function 

\begin{equation}
     \frac{d f_i}{d v_i} = \frac{d b(S)}{d v_i} - \frac{d c(v_i)}{d v_i} \,. \label{eqn:fitness_slope} 
\end{equation}

Since $\frac{db}{dv_i} = \frac{db}{dS}$ is the same for all group members, individual differ only in their marginal costs. When costs are steepening (convex), individuals with lower vigilance have lower marginal costs and are more incentivised to increase their vigilance, driving vigilances together. When costs are flattening (concave), the individual with higher vigilance has the lower marginal cost, driving vigilances apart.  

\begin{figure}
\centering
\includegraphics[width=0.8\linewidth]{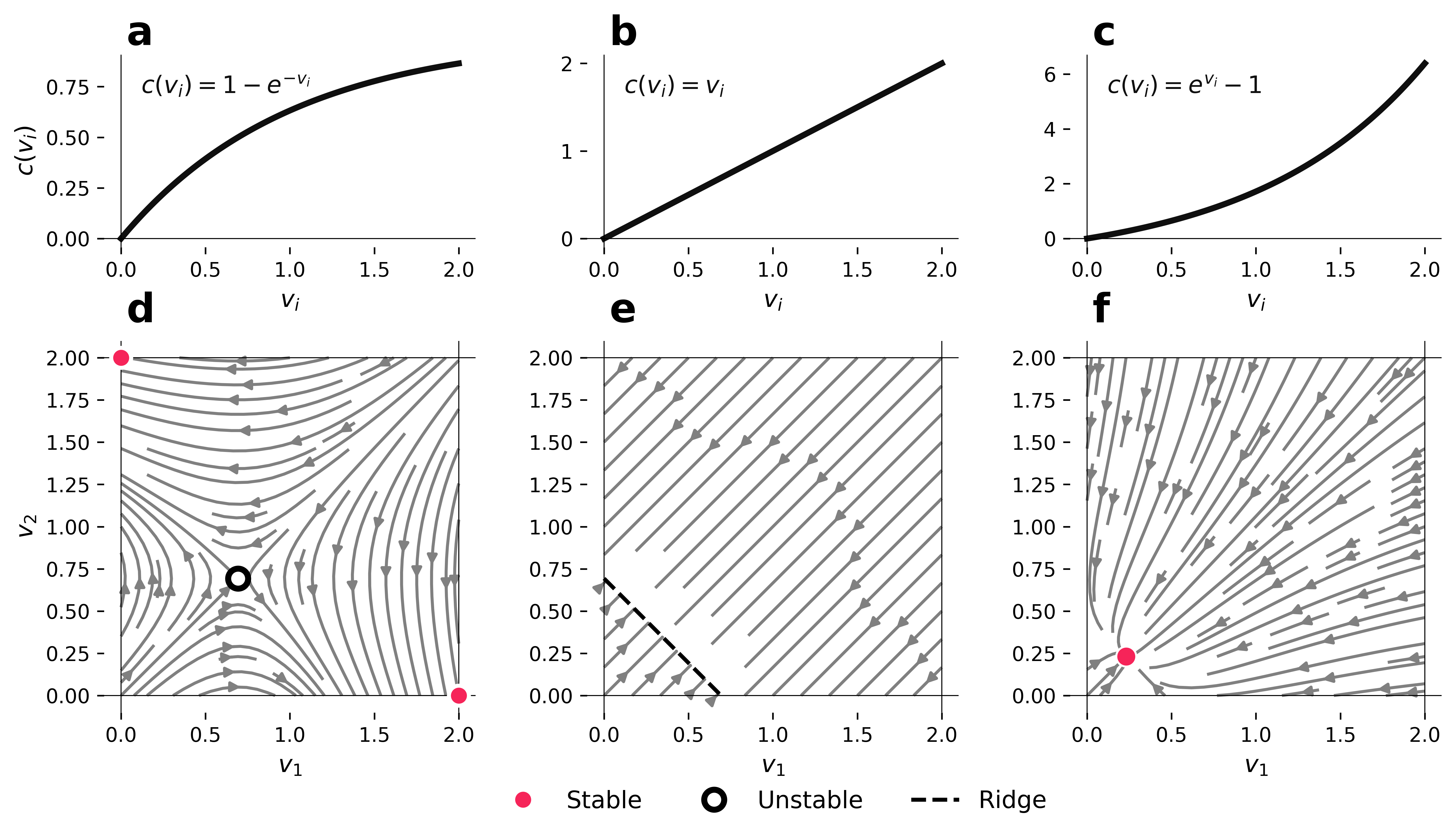}
\caption{Phase portraits of fitness gradients with a selfish group of size $N=2$, with individual vigilances of $v_1$ (x-axis) and $v_2$ (y-axis). Flow lines show how the group strategy changes. a) With flattening (concave) costs there is d) a single dynamically unstable interior equilibrium and $N$ symmetric stable equilibria on the edges of the strategy space. b) With linear vigilance costs there is e) a $N-1$ dimensional ridge of neutrally stable equilibria with a fixed optimal collective vigilance. c) With  steepening (convex) costs there is f) a single interior many-eyes stable point with all $N$ watchers having the same individual vigilance.}
\label{fig:dynamics_selfish}
\end{figure}

We generalise this argument with a dynamical systems analysis \cite{strogatz2001nonlinear}. Assuming that individuals adjust their vigilance in response to others, the system evolves according to the fitness gradients. Figure \ref{fig:dynamics_selfish} shows phase portraits for a group of $N=2$, with the flow lines showing how the group strategy is expected to change as individuals move towards higher fitness. The form of stable strategies is determined by the eigenvalues of the Jacobian matrix (see SI for full analysis), which in turn depend on the shape of the vigilance cost function: 

\begin{itemize}
    \item Flattening (concave) costs (Figure \ref{fig:dynamics_selfish} \textbf{a}, \textbf{d}): all interior points are unstable so that only edge (sentinel) solutions are stable. Individuals diverge to either zero or maximum vigilance, with at most one individual at an intermediate vigilance.    
    \item Linear costs (Figure \ref{fig:dynamics_selfish} \textbf{b}, \textbf{e}): interior solutions form a neutrally stable ridge. Strategies converge towards the optimal collective vigilance $S^*$, with any distribution of vigilances being stable.  
    \item Steepening (convex) costs (Figure \ref{fig:dynamics_selfish} \textbf{c}, \textbf{f}): interior (many-eyes) solutions are stable. All individuals converge to the same level of vigilance.        
\end{itemize} 

For completeness we investigated other solution concepts (see SI). A Nash equilibria analysis broadly agrees with the dynamical systems analysis. For convex (steepening) costs only many-eyes strategies are Nash equilibria. For sufficiently concave (flattening) costs, only sentinel strategies are Nash equilibria. However, if costs are concave but benefits are more concave, there exists both many-eyes and sentinel Nash equilibria. An evolutionarily stable strategy (ESS) analysis finds that only many-eyes strategies are evolutionarily stable; sentinel strategies are not stable since mutants with low vigilance can always free-ride and invade sentinel populations. These concepts address different biological questions: Nash equilibria describe static best responses, and ESS analysis is suitable when investigating how genetic traits invade populations (or resist invasion) \cite{smith1973logic, smith1982evolution}. Empirically, vigilance behaviour is flexible as individuals adjust their vigilance in response to group composition and the presence of sentinels \cite{elgar1989predator, manser1999response, bell2010bargaining, clutton1999selfish}. For this reason dynamical systems analysis, which captures ongoing behavioural adjustments, provides the most appropriate approach. 

\subsection{Cooperative Groups}

Cooperative groups are made up of $N$ individuals who are each aiming to maximise the average expected fitness of the group members \cite{mcnamara1992evolutionarily}. The optimal strategy is to distribute vigilance in the most efficient way. 

When costs are steepening (convex), it is most efficient to exploit the low marginal costs at low vigilance by distributing vigilance throughout the group in a many-eyes strategy: if Alice has higher vigilance than Bob then the average cost would be lower if Alice reduced her vigilance slightly and Bob increased his by the same amount. When costs are flattening (concave; environments with a vantage point), those with high vigilance have lower marginal costs so it is more efficient to concentrate vigilance in as few sentinels as possible. 

Formally, the average fitness is given by 

\begin{equation}
    \bar{f} = b(S) - \bar{c(v_i)} \,,
\end{equation}

where $\bar{f} = \frac{\sum_i f_i}{N}$ and $\bar{c(v_i)} = \frac{\sum_i c(v_i)}{N}$. For any fixed collective vigilance, $S=S^*$, the benefit $b(S^*)$ is fixed. In this case the distribution of vigilance $\textbf{v} = [v_1, v_2, ...,v_N]$ that maximises the average fitness in that which minimises the average costs, 

\begin{equation}
    \max_{\textbf{v}} \bar{f} = \min_{\textbf{v}} \bar{c} \\,   \quad S = S^*
\end{equation}

\begin{figure}
\centering
\includegraphics[width=1\linewidth]{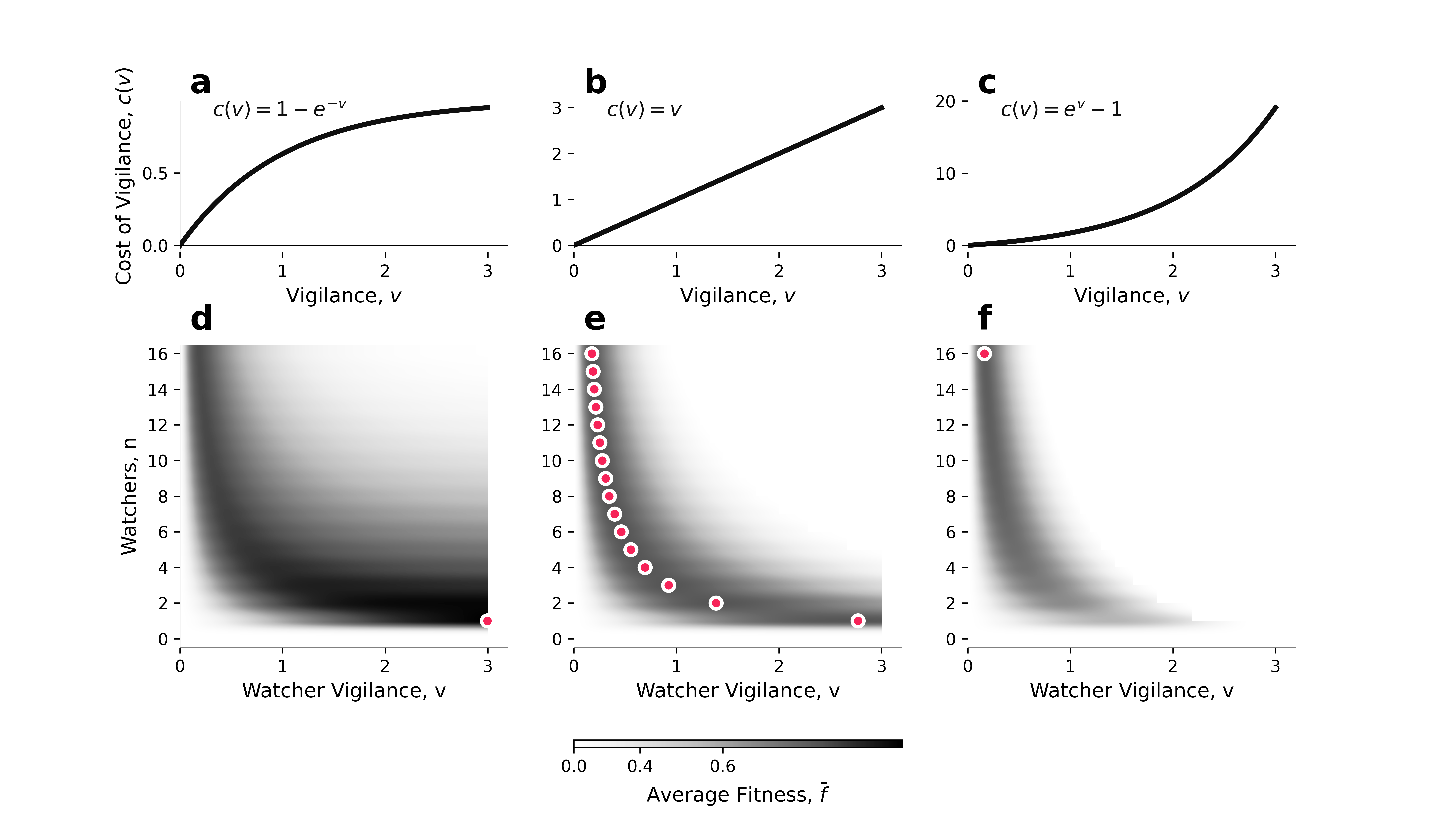}

\caption{Cooperative groups with $N=16$ individuals. The vigilance cost functions (a-c) determine the fitness landscapes panels d-f) over the number of watchers, n, with non-zero vigilance, and the vigilance of these watchers, v. Optimal strategies are denoted with black dots. a) Flattening (concave) vigilance costs favour d) a sentinel strategy with a single watcher, $n=1$, with high vigilance. b) Linear vigilance cost functions lead to e) indifference between strategies along an optimal ridge. c) Steepening (convex) vigilance costs favour f) a many-eyes strategy with the entire group, $n=16$, maintaining a low individual vigilance. Heatmaps only show positive average fitness, otherwise white.}
\label{fig:optimals}
\end{figure}

The optimal strategy is determined by the curvature of the vigilance cost function:

\begin{itemize}
    \item Flattening (concave) costs (Figure \ref{fig:optimals} \textbf{a}, \textbf{d}): Concentrating vigilance in sentinels minimises costs. If there is no maximum vigilance then a single sentinel is optimal, otherwise multiple sentinels can be preferred. 
    \item Linear costs (Figure \ref{fig:optimals} \textbf{b}, \textbf{e}): The group is indifferent because all distributions of vigilance for a given level of collective vigilance, $S^*$, have the same average cost.
    \item Steepening (convex) costs (Figure \ref{fig:optimals} \textbf{c}, \textbf{f}): Distributing vigilance equally across the many-eyes of the group minimises average cost and maximises fitness.    
\end{itemize}

If vigilance is concentrated in a single sentinel then the average cost of vigilance across the group is $\frac{c(S^*)}{N}$. If the same collective vigilance is spread equally through the many-eyes of the group, then the average cost of vigilance is $c(\frac{S^*}{N})$. Comparing these two averages, we can see (by Jensen's Inequality) that which is greater depends on the convexity/concavity of the costs of vigilance: when vigilance costs are steepening (convex) then the many-eyes strategy has lower costs; when vigilance costs are flattening (concave) then the sentinel strategy has lower average costs. See the SI for a general proof. 

Crucially, this result holds for any level of collective vigilance: the question of how to distribute vigilance (a many-eyes or sentinel strategy) is separate from the question of the optimal level of collective vigilance.

\subsection{Simulations}

\begin{figure}
\centering
\includegraphics[width=1\linewidth]{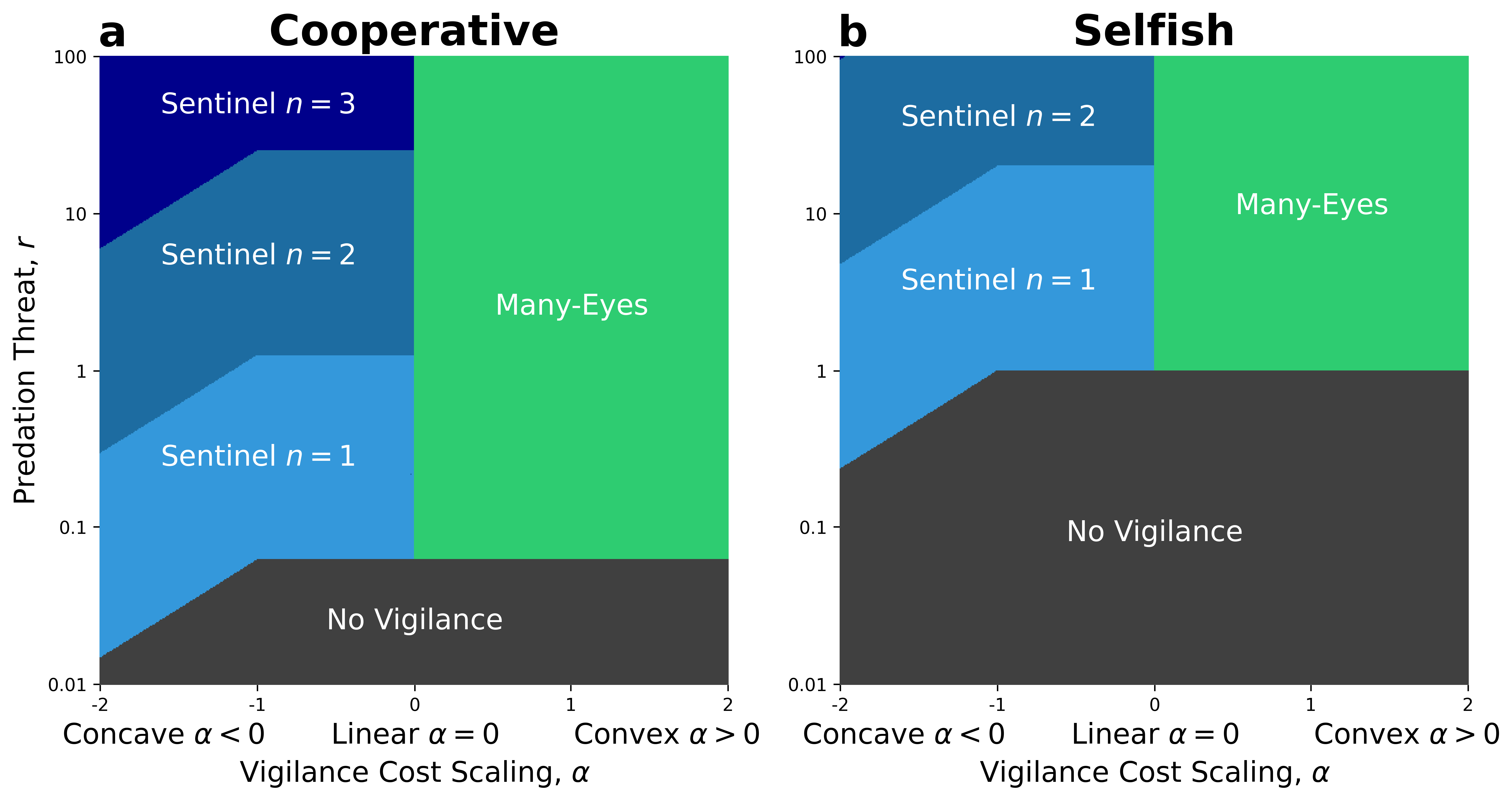}
\caption{Simulated vigilance behaviour in a) selfish and b) cooperative groups with varying scaling in vigilance costs (x-axis) and predation threat levels (y-axis log scale). In both selfish and cooperative groups, there is a clear dichotomy where we see many-eyes behaviour (green) with steepening (convex) costs and sentinel behaviour (blue) with flattening (concave) costs.}
\label{fig:heatmaps}
\end{figure}

We verify the general analysis through simulations with varying social context (selfish or cooperative), shapes of vigilance cost curves, and predation threat level, $r$ (Figure \ref{fig:heatmaps}). 

For each of the selfish and cooperative contexts, we ran a simulation for each point in a 500 x 500 parameter space described by the convexity/concavity, $\alpha \in [-2,2]$ (concave for $\alpha < 0$ and convex for $\alpha > 0$) and predation threat level $r \in [0.01, 100]$. Each simulation was initialised with a group of $N=16$ individuals, each with a uniformly randomly generated vigilance $v_i \in [0, v_{max}=10]$. At each timestep, an individual within the group was randomly selected to modify their vigilance to selfishly maximise their own fitness (Figure \ref{fig:heatmaps} \textbf{a}) or to cooperatively maximise the average group fitness (Figure \ref{fig:heatmaps} \textbf{b}), given the rest of the group's strategy. New vigilances were found using a scalar minimization algorithm over the interval $v_i \in [0, v_{max}=10]$ (using the python package scipy \cite{virtanen2020scipy}). Each simulation was run until the group vigilances converged. Note that we set a maximum vigilance, $v_{max}$, which leads to strategies with more than one sentinel as predation threat increases. Full details of the benefit and cost functions are in Materials and Methods. 

The simulations match the predictions from the analysis: many-eyes strategies appear when vigilance costs are steepening (convex) and sentinel strategies when vigilance costs are flattening (concave). Throughout the parameter space the cooperative groups have higher (or equal) average fitness than selfish groups. 

\subsection{Extensions}

Our analysis so far has been restricted to groups made up of identical individuals with relatively simple cost functions. Here, we discuss how the model can be extended towards more realistic scenarios. 

\begin{figure}
\centering
\includegraphics[width=1\linewidth]{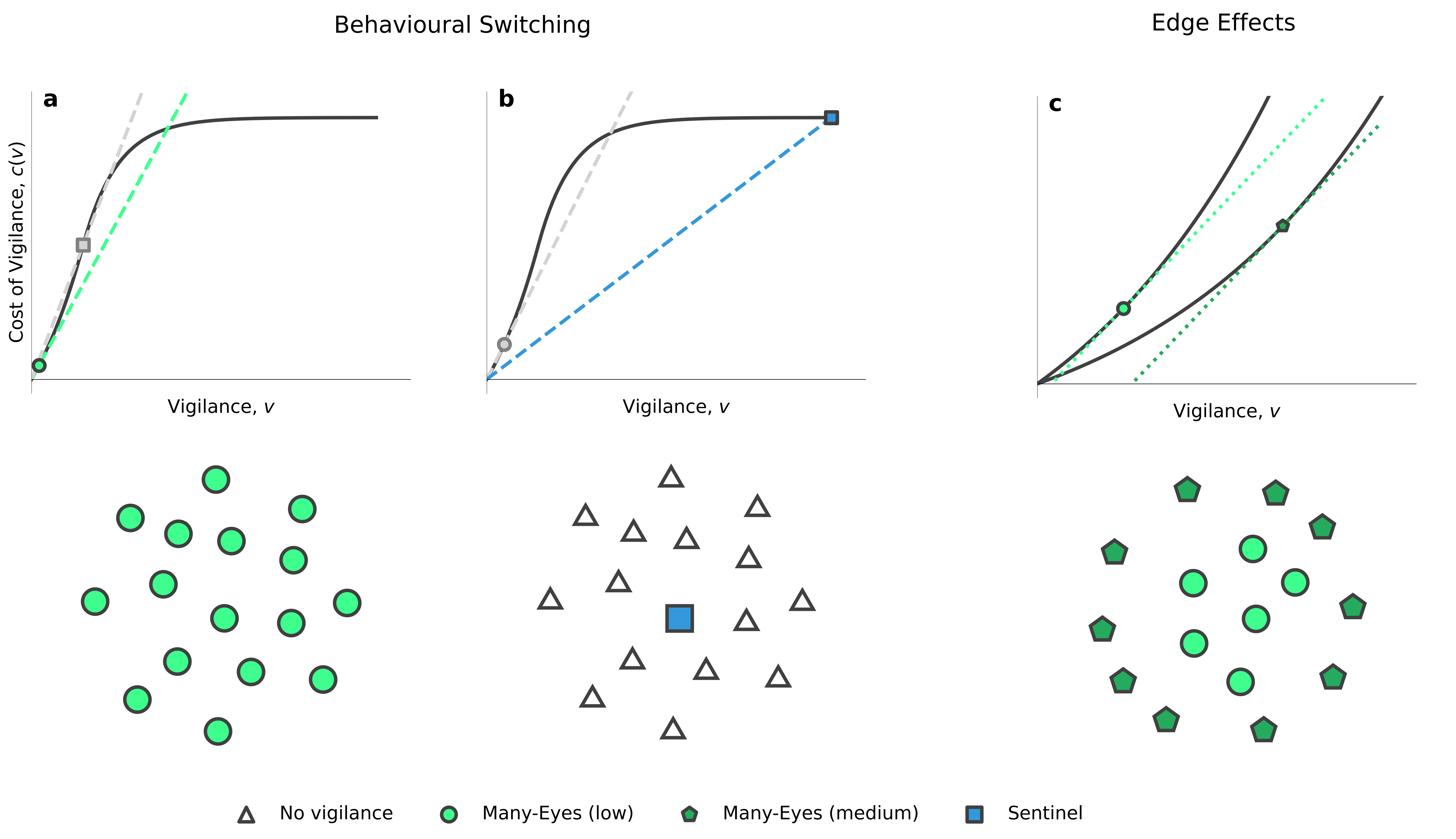}
\caption{Model extensions. a,b) Example of behavioural switching with S-shaped cost curves. a) At low predation threat, the many-eyes strategy (green circles) has a lower average vigilance cost (gradient of green dashed line) than a sentinel strategy (gradient of dashed grey line). b) At high predation threat levels, a sentinel strategy (blue square) has a lower average vigilance cost (gradient of dashed blue line) than a many-eyes strategy (gradient of dashed grey line). c) Example of edge effects with heterogenous cost curves. Lower costs at the edge of the group leads to higher vigilance at the edge (dark green pentagons) than the interior (green circles), with all individuals having the same marginal costs (dotted tangents).}
\label{fig:extensions}
\end{figure}

While environments with vantage points have flattening (concave) regions at high vigilance, it is reasonable to also expect a region of steepening (convex) costs at low vigilance (those foraging on the ground have the option to keep a low level lookout, even when visibility is low), forming a sigmoidal (S-shaped) vigilance cost function. These functions can support either many-eyes or sentinel behaviour, with each being locally optimal in the convex and concave regions respectively. Which is globally optimal depends on the full environmental conditions. In Figure \ref{fig:extensions} we show an example of a sigmoidal cost function. We simulated the optimal strategy for cooperative groups in the same way as described in the Simulations section. At low predation risk, the optimal level of collective vigilance is relatively low, so that it is more efficient to operate in the convex region of the curve, spreading the vigilance in a many-eyes strategy to exploit the low marginal costs at low vigilance (Figure \ref{fig:extensions} \textbf{a}). At high predation risk the optimal collective vigilance is greater, such that a many-eyes strategy now becomes less efficient and the group finds it worthwhile to switch to concentrate vigilance in a sentinel strategy, now in the concave region of the vigilance cost curve (Figure \ref{fig:extensions} \textbf{b}). The example given is for cooperative groups; we also simulated behavioural switching in selfish groups (which occurs at a higher predation threat rate). This simulated behavioural switching qualitatively matches empirical associations between the prevalence of sentinels and levels of predation threat \cite{bednekoff2015sentinel, mcgowan1989sentinel, ridley2010experimental}, and behavioural switching in meerkats such that the vigilance of foragers drops when sentinels are in position \cite{clutton1999selfish}. 

Real groups have heterogeneity due to spatial positions \cite{hoogland1979effect, lima1990behavioral}, energetic state \cite{clutton1999selfish, wright2001safe}, genetic variation \cite{olson2015exploring}, mixed species groups \cite{van2022modeling, barnard1982time}. In cooperative groups, heterogeneity in benefits will have no qualitative effect, as the group will still optimise the averaged benefit function. In selfish groups, heterogeneity in benefits are mathematically equivalent to heterogeneity in costs (as adaptive costs are relative to benefits). 

With heterogenous steepening (convex) vigilance costs, our model predicts many-eyes strategies with both cooperative and selfish groups, with differential levels of vigilance such that all individuals share the same marginal cost. Those with lower relative costs (e.g. peripheral individuals facing higher predation risk) maintain higher vigilance (Figure \ref{fig:extensions} \textbf{c}). These kinds of edge effects have been observed empirically in starlings \cite{jennings1980influence} and brent geese \cite{inglis1981vigilance}.  

With heterogenous flattening (concave) vigilance costs, we still see sentinel strategies. For cooperative groups, the optimal sentinel will be the individual with the lowest cost of vigilance. For selfish groups, there can be multiple individuals who could serve as sentinel in a stable equilibrium, but others who would not. For example, those who are satiated may find it beneficial to act as sentinel \cite{clutton1999selfish}. 

Our mean-field model assumes perfect information sharing. In reality, first observers can be more likely to escape a predator approach  \cite{bednekoff1997mutualism, bednekoff2001coordination, mcnamara1992evolutionarily, sirot2009coordination, bednekoff1998re, lima1990behavioral}. This can be incorporated in our model by combining private benefits with vigilance costs into a net cost function. The convexity principle still applies to the net cost function. This means that many-eyes behaviour is preferred when vigilance costs are more convex than private benefits (to generate a convex net cost function), otherwise a sentinel strategy is preferred.  

More complicated network effects where information spreads through neighbours can be addressed through spatially explicit models \cite{lima1996anti, sirot2009coordination}. In this case, we still expect a many-eyes/sentinel dichotomy to emerge within subgroups in the network. 

Finally, we do not explicitly include costs of coordination in our model. Some authors define the difference between many-eyes and sentinel behaviour as the degree of coordination \cite{bednekoff2015sentinel, bednekoff1997mutualism, mcgowan1989sentinel}, and we see adaptations to manage coordination such as specific calls when acting as sentinel \cite{bednekoff2015sentinel, wickler1985coordination, manser1999response}. Coordination costs could be included in our model either as an additional term or incorporated within the vigilance cost function. 

\section{Discussion}

Previous theoretical work on collective vigilance focusses on specific species, behaviours, or social contexts \cite{Pulliam1973Feb, pulliam1982scanning, mcnamara1992evolutionarily, bednekoff1997mutualism, bednekoff2001coordination}, leaving what Bednekoff \cite{bednekoff2015sentinel} calls ``large gaps in knowledge" in the conditions that lead to sentinel or many-eyes behaviour. We fill that gap with a general model that reveals a strict dichotomy: sentinel behaviour is favoured when there is an exploitable vantage point such that the costs of increasing vigilance are flattening (concave) and many-eyes when there is no vantage point and costs are steepening (convex). Notably, this same dichotomy holds in both selfish and cooperative groups. Our general approach synthesises disparate models in the literature and provides a framework for understanding collective vigilance behaviour across a wide variety of species and contexts. 

We extend our general model to more realistic sigmoidal vigilance costs, which we show can generate behavioural switching from many-eyes to sentinel strategies as predation threat rises. This theoretical finding aligns with empirical observations that sentinel behaviour occurs more often when predation threats are high \cite{bednekoff2015sentinel} as observed in Florida scrubjays \cite{mcgowan1989sentinel} and pied babblers \cite{ridley2010experimental}; as well as the discrete behavioural switching between many-eyes and sentinel strategies seen in meerkats \cite{clutton1999selfish}. We also extend our model to show how heterogeneity in the costs (and benefits) of vigilance can generate empirically observed edge effects where individuals on the edge of a group are more vigilant than those in the interior \cite{elgar1989predator, inglis1981vigilance, jennings1980influence}. These extensions demonstrate how our general framework can be used as a platform to provide a unified theoretical basis for a wide range of empirical collective vigilance behaviour. 

Our model predicts that sentinel behaviour will only emerge in ecological contexts where at least part of the cost function is flattening (concave), i.e. a behavioural option with steep followed by shallower marginal costs. For example, climbing a tree is costly but once up there it costs little extra to maintain a dedicated lookout. This matches the empirical pattern of sentinels occurring in species that forage on the ground in an area of low visibility for predator approaches but that have the availability of vantage points away from the ground with much better visibility \cite{bednekoff2015sentinel, wickler1985coordination, clutton1999selfish, gaston1977social, rasa1986coordinated, Horrocks_1986, bednekoff2001coordination, olson2015exploring}. However, the availability of a vantage point does not directly imply sentinel behaviour, and species must also have adaptations to be able to exploit that vantage point (e.g. perching in babblers, bipedal stances in meerkats, behavioural adaptations to coordinate vigilance). In the context of our model, one can say that vigilance costs depend on both the environment and species' traits. For example, many species of babblers (the genus \emph{Turdoides}) live in habitats with vantage points and exhibit sentinel behaviour \cite{gaston1977social, wright2001safe, wickler1985coordination, ridley2010experimental} even in habitats where other birds do not \cite{bednekoff2015sentinel}. We expect this pattern to hold more generally, and we predict that cross-species analyses of sentinel behaviour will reveal a strong association with the availability of vantage points in evolutionary habitats as well as clustering within closely related species with common sentinel ancestors.

Our model is a synthesis of decades of insightful empirical and theoretical work. However there are fundamental limitations on empirical work that may bias the behavioural patterns that we are able to observe. Vigilance is fundamentally difficult to measure \cite{elgar1989predator, beauchamp2015animal}. Much theoretical and empirical work on many-eyes is quantified by the raising of the head to scan for predators \cite{Pulliam1973Feb, pulliam1982scanning, hart1984vigilance, elgar1989predator, beauchamp2015animal}, but species can have multiple vigilance behaviours including maintaining a low level vigilance while their head is down \cite{lima1999back, jones2007vigilance, elgar1989predator, beauchamp2015animal}. Sentinels are associated with vantage points \cite{bednekoff2015sentinel, wickler1985coordination}, but this raises issues in that individuals on vantage points will be easier to observe by empiricists, it is not always clear whether an individual on a vantage point is acting as a sentinel for predators \cite{elgar1989predator, beauchamp2022visual, bednekoff2015sentinel}, and in any case we do not have a precise definition of a vantage point. These issues are limitations not only to our model but to the wider body of research in collective vigilance; our framework points to the importance of developing alternative empirical methods to quantify vigilance and its costs, and we suggest that a useful definition of a vantage point is the availability of a behavioural option with flattening (concave) vigilance costs. 

The model here is focussed on animal groups. However, our assumptions are very general and we hypothesise that our findings will also apply to other sensory systems. While speculative, we note what appear to be similar many-eyes and sentinel patterns in other domains. In human society, we see social vigilance structures that resemble many-eyes (neighbourhood watch \cite{bennett2006does}, public vigilance campaigns on public transport \cite{pearce2020encouraging}) and sentinel (lifeguards, security guards) behaviours. In technological systems, we see approaches with many low-cost sensors (distributed sensor networks \cite{brennan2004radiation}, cybersecurity \cite{mukherjee1994network}) and singular high-cost sensors (space telescopes, radar installations). And in biological morphology, we see many-eyes style sensory arrays (compound eyes \cite{land2012animal}, touch) and singular complex sensory organs (camera eyes \cite{land2012animal}, inner ears). While these domains have their own constraints and complexities, the generality of our model points towards a universal many-eyes/sentinel dichotomy in sensory systems. 

\section{Materials and Methods}

Our analysis is general and does not depend on specific forms for the benefit or cost functions, beyond the very general assumptions laid out in the Model section. However, we required specific forms in order to generate explanatory figures and simulations. 

Following common forms in the literature \cite{Pulliam1973Feb, pulliam1982scanning, delm1990vigilance}, for the benefit of collective vigilance we used 

\begin{equation}
    b(S) = r (1 - e^{-S}) \,. \label{eqn:benefit}
\end{equation}

This meets our conditions of increasing but saturating benefits of collective vigilance, and zero benefit at zero collective vigilance. 

For vigilance cost functions we used the general form 

\begin{equation}
    c(v) = \frac{e^{\alpha v} - 1}{\alpha} \,. 
\end{equation}

This meets our general conditions of an increasing cost function, zero cost at zero vigilance, and parameterises the convexity/concavity of the cost function with $\alpha < 0$ being concave and $\alpha > 0$ being convex. We chose this function in part because it has a similar derivative as the benefit cost function, so that the interior equilibrium of the fitness function is solvable analytically. 

The cost function reduces to linear costs as $\alpha \to 0$. To show this, we first note that the Taylor series of $e^x$ around $x=0$ is $e^x = \sum_{k=0} \frac{x^k}{k!}$. We can write 

\begin{equation}
    \lim_{\alpha \to 0} c(v) = \frac{(1 + \alpha v + \frac{(\alpha v)^2}{2!} + ...) - 1}{\alpha} \,,
\end{equation}

which simplifies to

\begin{equation}
    \lim_{\alpha \to 0} c(v) = v +  \frac{\alpha v^2}{2!} + \sim O(\alpha^2) \,,
\end{equation}

which converges to $\lim_{\alpha \to 0} c(v) = v$. 

\section{Data Availability}

All code and data is available at \url{https://github.com/chasmani/PUBLIC_many_eyes_and_sentinels_in_selfish_and_cooperative_groups}. 

\section{Acknowledgements}

This research was funded by a grant from the Templeton World Charity Foundation Inc. (TWCF-2021-20647) to D.B., R.P.M., and C.K. C.K. was also supported by the CIFAR Azrieli Global Scholars program.

\bibliographystyle{plain}
\bibliography{refs}

\section{Supplementary Information - Detailed Analysis for Selfish Groups}

\subsection{Dynamical Analysis}

We assume that individuals react to the vigilance strategies of others in order to increase their own fitness, i.e. the system will move down the fitness gradient given by $\mathbf{f'} = [\frac{d f_1}{d v_1}, \frac{d f_2}{d v_2}, ..., \frac{d f_N}{d v_N}]$. Where all individuals have zero fitness gradient there is an equilibrium point. We are interested in whether these equilibrium points are stable or unstable. 

To analyse stability we follow a standard dynamical systems analysis (see \cite{strogatz2001nonlinear}). One can linearise the dynamics around an equilibrium point through a Taylor expansion. This approximation captures how small perturbations develop through the Jacobian matrix, which contains the second derivatives of the fitness functions. The eigenvalues of the Jacobian determine whether perturbations grow or decay: if all eigenvalues are negative then perturbations decay and the equilibrium is stable; if any eigenvalue is positive then perturbations grow and the equilibrium is unstable. 

As described in the main paper, each individual has a fitness function 

\begin{equation}
    f_i = b(S) - c(v_i) \,.
\end{equation}

The Jacobian has entries 

\begin{equation}
    J_{ij} = \frac{d^2 f_i}{d v_i dv_j} \,.
\end{equation}

The off diagonal entries ($i \neq j$) are

\begin{equation}
    \frac{d^2 f_i}{d v_i d v_j} = \frac{d^2 b(S)}{d S^2} \,.
\end{equation}

Note that by the chain rule $\frac{d^2 b(S)}{d v_i v_j} = \frac{d^2 b(S)}{d S^2} \frac{d S}{dv_i} \frac{d S}{dv_j}$ and $\frac{d S}{dv_i} = \frac{d S}{dv_j} = 1$, so that $\frac{d^2 b(S)}{d v_i v_j} = \frac{d^2 b(S)}{d S^2}$. 

The diagonal entries ($i = j$) of the Jacobian are

\begin{equation}
    \frac{d^2 f_i}{d v_i^2} = \frac{d^2 b(S)}{d S^2} - \frac{d^2 c(v_i)}{d v_i^2} \,.
\end{equation}

The Jacobian can therefore be written as 

\begin{equation}
    J = \begin{pmatrix}
\alpha - \beta & \alpha & \alpha & \cdots & \alpha \\
\alpha & \alpha - \beta & \alpha & \cdots & \alpha \\
\alpha & \alpha & \alpha - \beta & \cdots & \alpha \\
\vdots & \vdots & \vdots & \ddots & \vdots \\
\alpha & \alpha & \alpha & \cdots & \alpha - \beta
\end{pmatrix} \,,
\end{equation}

where $\alpha = \frac{d^2 b(S)}{d S^2}$ and $\beta = \frac{d^2 c(v)}{d v^2}$. Note that at an interior equilibrium point $\frac{d^2 c(v_i)}{d v_i^2}$ is the same for all $i$. 

We can write this compactly as 

\begin{equation}
    J = \alpha \mathbf{1}\mathbf{1}^T - \beta I   \,, \label{eqn:jacobian_compact}
\end{equation}

where and $I$ is the identity matrix and $\mathbf{1}^T = [1,1,...,1]$ and therefore $\mathbf{1}\mathbf{1}^T$ is a matrix with all entries equal to 1. 

An eigenvalue, $\lambda_k$, and associated eigenvector, $\mathbf{u_k}$, is defined by the condition $J \mathbf{u_k} = \lambda_k \mathbf{u_k}$, i.e.,

\begin{equation}
    (\alpha \mathbf{1}\mathbf{1}^T - \beta I) \mathbf{u_k} = \lambda_k \mathbf{u_k} \,. \label{eqn:eigen_condition}
\end{equation}

We assume that an eigenvector is the vector $\mathbf{u_1} = \mathbf{1}$. Substituting into Equation \ref{eqn:eigen_condition}, 

\begin{equation}
    \alpha \mathbf{1}\mathbf{1}^T\mathbf{1} - \beta \mathbf{1} = \lambda_k \mathbf{1}\,.
\end{equation}

The term $\mathbf{1}^T\mathbf{1} = N$, so that we can write 

\begin{equation}
    (N \alpha - \beta) \mathbf{1} = \lambda_k \mathbf{1}\,.
\end{equation}

This fulfils the eigenvector condition. Therefore, $\mathbf{u_1} = \mathbf{1}$ is an eigenvector with eigenvalue $\lambda_1 = N \alpha - \beta$. 

The remaining eigenvectors form a basis such that all are orthogonal to $\mathbf{u_1}$. To see this, assume an eigenvector, $\mathbf{u_2}$, orthogonal to the first eigenvector, so that $\mathbf{1}^T \mathbf{u_2} = 0$. (such as $\mathbf{u_2} = [1,-1,0,0,...,0]$). Substituting this eigenvector into Equation \ref{eqn:eigen_condition}, 

\begin{equation}
    \alpha \mathbf{1}\mathbf{1}^T\mathbf{u_2} - \beta \mathbf{u_2} = \lambda_k \mathbf{u_2} \,.
\end{equation}

By the orthogonal definition $\mathbf{1}^T \mathbf{u_2} = 0$, so the first term is equal to zero and we can write

\begin{equation}
    - \beta \mathbf{u_2} = \lambda_k \mathbf{u_2} \,.
\end{equation}

This fulfills the eigenvector condition, so that any off-diagonal vector (orthogonal to $\mathbf{1}$) is an eigenvector with eigenvalue $\lambda_2 = - \beta$. 

If we consider a basis that includes $\mathbf{u_1} = \mathbf{1}$, we will find that there are $N-1$ vectors orthogonal to $\mathbf{1}$ (with $N$ individuals the strategy space is $N$-dimensional), each of which is an eigenvector with eigenvalue $\lambda = - \beta$. This gives us the eigenvalues

\begin{equation}
    \lambda_k = \begin{cases}
        N\frac{d^2 b}{dS^2} - \frac{d^2c}{dv^2} & k = 1 \\
        -\frac{d^2c}{dv^2} & k = 2, 3, ..., N
    \end{cases} \,.
\end{equation}

The eigenvalues of the Jacobian determine the stability of interior equilibria. An equilibrium is stable if \emph{all} eigenvalues are negative, $\lambda_k < 0 \quad \forall \quad k$. Conversely, an equilibrium is unstable if \emph{any} eigenvalue is positive, $\exists k : \lambda_k > 0$. 

If costs are steepening (convex), $\frac{d^2 c}{dv^2} > 0$, then the off-diagonal eigenvectors are negative. In the main text, we defined the benefit function as concave because benefits saturate, $\frac{d^2 b}{dS^2} < 0$, so that the first (diagonal) eigenvalue is also negative. Therefore, all eigenvalues are negative and an interior equilibrium is stable. A many-eyes strategy involves all (or many) individuals having low vigilance, $0 < v < v_{max}$, which represents an interior equilibrium. Therefore, a many-eyes strategy is stable if vigilance costs are steepening (convex).

If costs are flattening (concave), $\frac{d^2 c}{dv^2} < 0$, then the off-diagonal eigenvalues are positive, so an interior equilibrium is unstable. For dynamical instability, the sign of the first (diagonal) eigenvalue is not important because the condition is for any eigenvalue to be positive, which is fulfilled by the off-diagonal eigenvalues. Any strategy with more than 1 individual with an intermediate vigilance $0 < v < v_{max}$ represents an interior solution, and will therefore be unstable if vigilance costs are flattening (concave). Therefore, stable strategies can only exist at the edges of the strategy space, with at least $N-1$ individuals having either zero vigilance or maximum vigilance. These edge solutions correspond to a sentinel strategy with one (or a few) vigilant individuals and all others having zero vigilance.

\subsection{Nash Equilibria}

Nash equilibria are strategy distributions where no individual is incentivised to switch strategies. An individual $i$ has fitness

\begin{equation}
    f_i(v_i, \textbf{v}_{\sim i}) = b(v_i + S_{\sim i}) - c(v_i) \,,
\end{equation}

where $\textbf{v}_{\sim i}$ is the strategy vector excepting player $i$, $S_{\sim i}$ is the collective vigilance excepting player $i$, $b$ is the benefit of collective vigilance and $c$ is the individual cost of vigilance. We assume that benefits are increasing and concave. We assume that costs are increasing and are either convex, linear, or concave. 

For a Nash equilibrium, all individuals must be playing a best response to the strategy of others, i.e. they cannot improve their expected fitness by unilaterally changing their vigilance.

\subsubsection{Symmetric Many-Eyes Nash Equilibrium}

All individuals play the same $v^* \in (0, v_{max})$. A Nash equilibrium must fulfill the first order condition:

\begin{equation}
    \frac{d b(S)}{d S} \bigg|_{S=Nv^*}= \frac{d c(v)}{d v} \bigg|_{v=v^*}\,.
\end{equation}

Note that $\frac{db}{dS} = \frac{db}{dv}$. This is a Nash equilibrium if the second-order condition is fulfilled that the fitness function is locally non-convex, $\frac{d^2 f}{dv^2} \leq 0$, 

\begin{equation}
    \frac{d^2 b(S)}{d S^2} \bigg|_{S=Nv^*} \leq \frac{d^2 c(v)}{d v^2} \bigg|_{v=v^*} \,.
\end{equation}

Many-eyes Nash equilibria are possible when vigilance costs are more convex than benefits. 

\subsubsection{Zero Vigilance Nash Equilibrium}

All individuals have zero vigilance, $v^* = 0$. This is a Nash equilibrium if there is no level of vigilance that has a greater benefit than cost

\begin{equation}
    b(v) < c(v) \, \quad \forall v \,.
\end{equation}

In the main paper we defined the benefit and cost functions to start at zero, i.e. $b(0)=0$ and $c(0)$. The zero vigilance Nash equilibrium locally requires 

\begin{equation}
    \frac{d b(S)}{dS} \bigg|_{S=0} \leq \frac{dc(v)}{dv} \bigg|_{v=0} \,.
\end{equation}

As benefits (and marginal benefits) are scaled by the predation threat level, $r$, we can expect a zero vigilance Nash equilibrium when predation threat levels are low. 

\subsubsection{Maximum Vigilance Nash Equilibrium}

All individuals have maximum vigilance $v^* = v_{max}$. This is a Nash equilibrium if there is no vigilance $v \in [0, v_{max}]$ that can benefit from deviating, 

\begin{equation}
    b(N v_{max}) - c(v_{max}) \geq  b((N-1) v_{max} + v) - c(v) \,, \quad \forall v \,. 
\end{equation}

Locally, a Nash equilibrium requires 

\begin{equation}
    \frac{d b(S)}{d v} \bigg|_{S=N v_{max}} \geq \frac{d c(v)}{dv} \bigg|_{v=v_{max}} \,.
\end{equation}

This condition is strong as it requires that marginal benefits at the maximum collective vigilance of the entire group are greater than the marginal costs of each individual. 

\subsubsection{Non-symmetric Interior Nash Equilibria}

We define a non-symmetric interior Nash equilibrium as requiring at least two individuals with different vigilances $v_A, v_B \in (0, v_{max})$. 

They must both fulfill the first order condition at the optimal collective vigilance $S^*$,

\begin{equation}
    \frac{d b(S)}{dv} \bigg|_{S=S^*} = \frac{d c}{dv} \bigg|_{v=v_A, v=v_B} \,.
\end{equation}

The marginal benefit is the same for both individuals, as it is a function of the collective vigilance. In order for both individuals to be at equilibrium, they must have the same marginal costs. This is not possible with convex or concave vigilance cost functions. Non-symmetric Nash equilibria are possible if the vigilance cost function is linear. 

If the vigilance cost function is linear, then the fitness function is concave (because benefits are concave), and so the second order condition of a locally concave fitness function is fulfilled. One can determine the optimal collective vigilance by finding where the marginal benefit equals the linear cost gradient. Any distribution of vigilances that sum to the optimal collective vigilance is a Nash equilibrium. 

\subsubsection{Sentinel Nash Equilibria}

A sentinel Nash equilibrium has $n_{int}$ sentinels with an interior vigilance $v_{int} \in (0, v_{max})$, $n_{max}$ sentinels with maximum vigilance and $N-n_{max}-n_{int}$ non-sentinels with zero vigilance, such that $n_{max} + n_{int} \geq 1$.

The non-sentinels are at equilibrium if they cannot improve by deviating from zero vigilance

\begin{equation}
    b(S^* + v) \leq c(v) \,, \quad \forall v \,.
\end{equation}

Locally, they must have a non-positive fitness gradient

\begin{equation}
    \frac{d b(S)}{dv} \bigg|_{S=S^*}\leq \frac{d c(v)}{dv} \bigg|_{v=0} \,.
\end{equation}

The sentinels at maximum vigilance cannot deviate by decreasing their vigilance. Locally, 

\begin{equation}
    \frac{d b(S)}{dv} \bigg|_{S=S^*} \geq \frac{d c(v)}{dv} \bigg|_{v=v_{max}} \,.
\end{equation}

A sentinel with an interior vigilance must locally have zero fitness function 

\begin{equation}
    \frac{d b(S)}{dv} \bigg|_{S=S^*} = \frac{d c(v)}{dv} \bigg|_{v=v_{int}} \,.
\end{equation}

The marginal benefit is the same for all members of the group. The marginal costs must be higher at zero vigilance than maximum vigilance. This is possible if the cost function is concave or linear, 

\begin{equation}
    \frac{d^2 c}{dv^2} \leq 0 \,.
\end{equation}

\subsubsection{Summary}

The many-eyes and sentinel conditions are not equivalent, and we can describe three distinct regions:

\begin{itemize}
    \item $\frac{d^2 c}{dv^2} < \frac{d^2 b}{dv^2}$. Sentinel NE are possible but many-eyes are not. 
    \item $\frac{d^2 b}{dv^2} \leq \frac{d^2 c}{dv^2} \leq 0$. Both sentinel and many-eyes NE are possible. 
    \item $0 < \frac{d^2 c}{dv^2}$. Many-eyes NE are possible but sentinels are not.  
\end{itemize}

Note that dynamically stable equilibria are also Nash equilibria, but Nash equilibria are not necessarily dynamically stable. The Nash equilibrium analysis is consistent with the dynamical stability analysis in that many-eyes solutions are possible when $0 \leq \frac{d^2 c}{dv^2}$ and sentinel strategies are possible when $\frac{d^2 c}{dv^2} \leq 0$. However we find a region that allows both many-eyes and sentinel Nash equilibria in in $\frac{d^2 b}{dv^2} < \frac{d^2 c}{dv^2} < 0$, this suggests that sentinel strategies in this region are dynamically unstable Nash equilibria. 

\subsection{Evolutionarily Stable Strategies}

A common solution concept in the literature is evolutionarily stable strategies (ESS), which was an innovation led by Maynard-Smith \cite{smith1973logic, smith1982evolution} with the intention of applying game theory to evolutionary settings by restricting the dynamism through which populations could change strategies, to reflect the constraints of fixed genotypes within lifetimes. Evolutionarily stable strategies ask whether a mutant can invade a population with a fixed equilibrium strategy. 

Within our notation, a condition for ESS is \cite{smith1973logic}

\begin{equation}
    f(v_m, \textbf{v}) <  f(v', \textbf{v}) \,,
\end{equation}

where $v_m$ is the vigilance of a mutant, $v'$ is the vigilance of a resident, $\textbf{v}$ are the vigilances of the group. 

We begin with the fitness equation described in the main paper

\begin{equation}
    f_i = b(S) - c(v_i) \,, \quad S = \sum_i v_i \,,
\end{equation}

with the terms as defined in the main paper. We assume $\frac{d b}{d S} > 0$, $\frac{d^2 b}{d S^2} < 0$, $\frac{d c}{d v_i} > 0$. 

\subsubsection{Many-eyes ESS}

A many-eyes strategy has all individuals with the same vigilance, $v^*$, giving a total vigilance $S = N v^*$. 

An ESS must fulfill the first order (best-response) condition

\begin{equation}
    \frac{d b(S)}{dv} \bigg|_{S=Nv^*} = \frac{d c(v)}{dv} \bigg|_{v=v^*}\,.
\end{equation}

Now consider a mutant with higher vigilance $v_m > v^*$. The mutant has greater fitness if the increased benefits are greater than the increased costs, which depends on how the slopes change. If $\frac{d^2 b}{dv^2} < \frac{d^2 c}{dv^2}$ then the mutant's extra benefits are smaller than their extra costs, and the mutant cannot invade. 

The same logic applies for a mutant with lower vigilance than $v^*$. In this case, the mutant cannot invade if $\frac{d^2 b}{dv^2} < \frac{d^2 c}{dv^2}$ because benefits reduce more quickly than costs. 

Hence, a many-eyes strategy is evolutionarily stable if the vigilance costs are more convex than the benefits, 

\begin{equation}
    \frac{d^2 c}{dv^2} > \frac{d^2 b}{dv^2} \,.
\end{equation}

\subsubsection{Sentinel ESS}

We assume that a sentinel strategy has $n_s$ individuals with vigilance $v_s >0$ and all other individuals with zero vigilance, giving collective vigilance $S=v_s$. 

The sentinel's fitness is 

\begin{equation}
    f_s = b(n_s v_s) - c(v_s) \,.
\end{equation}

A non-sentinel's fitness is 

\begin{equation}
    f_i = b(n_s v_s) \,.
\end{equation}

Consider a mutant with vigilance $v_m$, 

\begin{equation}
    f_m = b(n_s v_s + v_m) - c(v_m) \,.
\end{equation}

Trivially, the mutant fitness is higher than the sentinel's fitness for $v < v_s$, because the mutant is receiving the full benefit of the sentinel's vigilance without matching their costs. Therefore, a mutant can invade and a sentinel strategy is not evolutionarily stable, given these assumptions. 

We would therefore predict that purely genetically determined vigilance behaviour in selfish groups should only generate many-eyes strategies, and we should not see sentinels in such situations. One could expand the ESS analysis to more complicated vigilance strategies such as ``\emph{act as sentinel if no-one else is and I am satiated}" \cite{bednekoff1997mutualism}. 

\section{Supplementary Information - Detailed Analysis for Cooperative Groups}

In the main article we assume that cooperative groups are incentivised to maximise the average fitness

\begin{equation}
    \bar{f} = b(S) - \bar{c(v_i)} \,,
\end{equation}

For a given collective vigilance, $S^*$, we can ask what distribution of group vigilances, $\mathbf{v} = [v_1, v_2, ..., v_N]$, maximises fitness. The benefit is fixed at $b(S^*)$. Maximising average fitness is therefore equivalent to minimising the average (or total) cost 

\begin{equation}
    \max_{\mathbf{v}} \bar{f}(S^*, \mathbf{v}) = \min_{\mathbf{v}} \frac{1}{N} \sum_{\mathbf{v}}c(v_i)\,, \quad S^* = \sum_i v_i \,.
\end{equation}

Consider a group of individuals, with two of the individuals having different vigilances $v_{high}$ and  $v_{low}$ such that $v_{high} \geq v_{low}$. These two individuals pay a total cost of $c(v_{high}) + c(v_{low})$. 

The group could achieve the same total vigilance if these two individuals instead had vigilances $v_{high}' = v_{high} + \Delta v$ and $v_{low}' = v_{low} - \Delta v$, where we label them such that $v_{high}' \geq v_{low}'$. Note that this labelling convention holds for any $\Delta v$, including cases where the individuals' relative positions switch. Crucially, this means $v_{high} > v'_{low}$. 

The change in cost paid following this switch is 

\begin{equation}
    \Delta c = c(v_{high}) + c(v_{low}) - c(v'_{high}) - c(v'_{low}) \,.
\end{equation}

We can write this as

\begin{equation}
    \Delta c = [c(v_{\text{high}}) - c(v_{\text{high}} + \Delta v)] - [c(v'_{\text{low}}) - c(v'_{\text{low}} + \Delta v)] \,.
\end{equation}

By our construction $v_{high} > v'_{low}$. We are therefore the comparing the change in costs of two equal increments in vigilance ($\Delta v)$ at two different baseline levels. the sign of $\Delta c$ depends on both the sign of $\Delta v$ and the convexity/concavity of the vigilance cost function:

\begin{itemize}
    \item Convex costs with diverging vigilance ($\Delta v > 0$). The cost increase at the higher baseline is greater than the cost decrease at the lower level. Therefore $\Delta c > 0$ (less efficient).
    \item Convex costs with converging vigilance ($\Delta v < 0$). The cost decrease at the higher baseline is greater than the cost increase at the low baseline. Therefore $\Delta c < 0$ (more efficient).
    \item Concave costs with diverging vigilance ($\Delta v > 0$). The cost increase at the higher baseline is smaller than the cost decrease at the lower baseline. Therefore $\Delta c < 0$ (more efficient).
    \item Concave costs with converging vigilance ($\Delta v < 0$). The cost decrease at the higher baseline is smaller than the cost increase at the lower baseline. Therefore $\Delta c > 0$ (less efficient).
\end{itemize}

Putting these outcomes together, we can conclude that for convex costs it is always more efficient to have more equal vigilances, so that the optimal group strategy is an equal distribution of vigilance, i.e. many-eyes. For concave costs, it is always more efficient for vigilance to diverge, so that the optimal group strategy is to have vigilances at zero or maximum vigilance, with at most one individual with an intermediate vigilance. 

This argument applies for any level of collective vigilance, and so is independent of $S$ and independent of the specific form of the predation cost rate, $b(S)$. In plain English, sentinels are preferred over many-eyes for concave cost functions, and vice-versa for convex cost functions, and which is preferred does not depend on the level of collective vigilance. 

In the concave case, if vigilance is able to increase indefinitely then it is optimal to have one individual with the full collective vigilance, $v_{sentinel} = S^*$, and everyone else with zero. However, where there is a maximum possible vigilance then it can be beneficial to have multiple sentinels. 

\end{document}